\documentclass[dvips]{article}
\usepackage{supertabular,lscape,epsfig}
\usepackage{amssymb}
\usepackage{amsmath}

\DeclareSymbolFont{ppa}{OT1}{ppl}{m}{it}
\DeclareMathSymbol{\vv}{\mathalpha}{ppa}{'166}

\begin{document}

\newcommand{\dd}{\,{\rm d}}
\newcommand{\ie}{{\it i.e.},\,}
\newcommand{\etal}{{\it et al.\ }}
\newcommand{\eg}{{\it e.g.},\,}
\newcommand{\cf}{{\it cf.\ }}
\newcommand{\vs}{{\it vs.\ }}
\newcommand{\zdot}{\makebox[0pt][l]{.}}
\newcommand{\up}[1]{\ifmmode^{\rm #1}\else$^{\rm #1}$\fi}
\newcommand{\dn}[1]{\ifmmode_{\rm #1}\else$_{\rm #1}$\fi}
\newcommand{\upd}{\up{d}}
\newcommand{\uph}{\up{h}}
\newcommand{\upm}{\up{m}}
\newcommand{\ups}{\up{s}}
\newcommand{\arcd}{\ifmmode^{\circ}\else$^{\circ}$\fi}
\newcommand{\arcm}{\ifmmode{'}\else$'$\fi}
\newcommand{\arcs}{\ifmmode{''}\else$''$\fi}
\newcommand{\MS}{{\rm M}\ifmmode_{\odot}\else$_{\odot}$\fi}
\newcommand{\RS}{{\rm R}\ifmmode_{\odot}\else$_{\odot}$\fi}
\newcommand{\LS}{{\rm L}\ifmmode_{\odot}\else$_{\odot}$\fi}

\newcommand{\Abstract}[2]{{\footnotesize\begin{center}ABSTRACT\end{center}
\vspace{1mm}\par#1\par
\noindent
{~}{\it #2}}}

\newcommand{\TabCap}[2]{\begin{center}\parbox[t]{#1}{\begin{center}
  \small {\spaceskip 2pt plus 1pt minus 1pt T a b l e}
  \refstepcounter{table}\thetable \\[2mm]
  \footnotesize #2 \end{center}}\end{center}}

\newcommand{\TableSep}[2]{\begin{table}[p]\vspace{#1}
\TabCap{#2}\end{table}}

\newcommand{\FigCap}[1]{\footnotesize\par\noindent Fig.\  %
  \refstepcounter{figure}\thefigure. #1\par}

\newcommand{\TableFont}{\footnotesize}
\newcommand{\TableFontIt}{\ttit}
\newcommand{\SetTableFont}[1]{\renewcommand{\TableFont}{#1}}

\newcommand{\MakeTable}[4]{\begin{table}[htb]\TabCap{#2}{#3}
  \begin{center} \TableFont \begin{tabular}{#1} #4 
  \end{tabular}\end{center}\end{table}}

\newcommand{\MakeTableSep}[4]{\begin{table}[p]\TabCap{#2}{#3}
  \begin{center} \TableFont \begin{tabular}{#1} #4 
  \end{tabular}\end{center}\end{table}}

\newenvironment{references}%
{
\footnotesize \frenchspacing
\renewcommand{\thesection}{}
\renewcommand{\in}{{\rm in }}
\renewcommand{\AA}{Astron.\ Astrophys.}
\newcommand{\AAS}{Astron.~Astrophys.~Suppl.~Ser.}
\newcommand{\ApJ}{Astrophys.\ J.}
\newcommand{\ApJS}{Astrophys.\ J.~Suppl.~Ser.}
\newcommand{\ApJL}{Astrophys.\ J.~Letters}
\newcommand{\AJ}{Astron.\ J.}
\newcommand{\IBVS}{IBVS}
\newcommand{\PASP}{P.A.S.P.}
\newcommand{\Acta}{Acta Astron.}
\newcommand{\MNRAS}{MNRAS}
\renewcommand{\and}{{\rm and }}
\section{{\rm REFERENCES}}
\sloppy \hyphenpenalty10000
\begin{list}{}{\leftmargin1cm\listparindent-1cm
\itemindent\listparindent\parsep0pt\itemsep0pt}}%
{\end{list}\vspace{2mm}}

\def\TYLDA{~}
\newlength{\DW}
\settowidth{\DW}{0}
\newcommand{\dw}{\hspace{\DW}}

\newcommand{\refitem}[5]{\item[]{#1} #2%
\def\REFARG{#3}\ifx\REFARG\TYLDA\else, {\it#3}\fi
\def\REFARG{#4}\ifx\REFARG\TYLDA\else, {\bf#4}\fi
\def\REFARG{#5}\ifx\REFARG\TYLDA\else, {#5}\fi.}

\newcommand{\Section}[1]{\section{\hskip-6mm.\hskip3mm#1}}
\newcommand{\Subsection}[1]{\subsection{#1}}
\newcommand{\Acknow}[1]{\par\vspace{5mm}{\bf Acknowledgements.} #1}
\pagestyle{myheadings}

\newfont{\bb}{ptmbi8t at 12pt}
\newcommand{\xrule}{\rule{0pt}{2.5ex}}
\newcommand{\xxrule}{\rule[-1.8ex]{0pt}{4.5ex}}
\def\thefootnote{\fnsymbol{footnote}}
\begin{center}
{\Large\bf The Optical Gravitational Lensing Experiment.
\vskip1pt
Planetary and Low-Luminosity Object Transits
\vskip1pt
in the Carina Fields of the Galactic Disk\footnote{Based on observations
obtained with the 1.3~m Warsaw telescope at the Las Campanas Observatory
of the Carnegie Institution of Washington.}}
\vskip.6cm
{\bf
A.~~U~d~a~l~s~k~i$^1$,~~O.~~S~z~e~w~c~z~y~k$^1$,~~K.~~\.Z~e~b~r~u~\'n$^1$,
~~G.~~P~i~e~t~r~z~y~\'n~s~k~i$^{2,1}$,~~M.~~S~z~y~m~a~{\'n}~s~k~i$^1$,
~~M.~~K~u~b~i~a~k$^1$, ~~I.~~S~o~s~z~y~\'n~s~k~i$^1$,
~~and~~\L.~~W~y~r~z~y~k~o~w~s~k~i$^1$}
\vskip2mm
$^1$Warsaw University Observatory, Al.~Ujazdowskie~4, 00-478~Warszawa, Poland\\
e-mail: (udalski,szewczyk,zebrun,pietrzyn,msz,mk,soszynsk,wyrzykow)@astrouw.edu.pl\\
$^2$ Universidad de Concepci{\'o}n, Departamento de Fisica,
Casilla 160--C, Concepci{\'o}n, Chile
\end{center}

\Abstract{We present results of the second ``planetary and low-luminosity 
object transit'' campaign conducted by the OGLE-III survey. Three fields 
($35\arcm\times35\arcm$ each) located in the Carina regions of the Galactic 
disk ($l\approx290\arcd$) were monitored continuously in February--May 2002. 
About 1150 epochs were collected for each field. 

The search for low depth transits was conducted on about 103~000 stars with 
photometry better than 15~mmag. In total, we discovered 62 objects with 
shallow depth ($\leq0.08$~mag) flat-bottomed transits. For each of these 
objects several individual transits were detected and photometric elements were 
determined. Also lower limits on radii of the primary and companion were 
calculated. 

The 2002 OGLE sample of stars with transiting companions contains
considerably  more objects that may be Jupiter-sized (${R<1.6~R_{\rm
Jup}}$) compared to  our 2001 sample. There is a group of planetary 
candidates with the orbital periods close to or shorter than one day. If
confirmed as  planets, they would be the shortest period extrasolar
planetary systems. 

In general, the transiting objects may be extrasolar planets, brown dwarfs, or 
M-type dwarfs. One should be, however, aware that in some cases unresolved 
blends of regular eclipsing stars can mimic transits. Future spectral analysis  
and eventual determination of the amplitude of radial velocity  should allow 
final classification. High resolution spectroscopic follow-up observations 
are, therefore, strongly encouraged. 

All photometric data are available to the astronomical community from the OGLE 
{\sc Internet} archive.}{}

\Section{Introduction} 
\vspace*{13pt}
Observations of photometric transits caused by small objects passing on the 
disk of a hosting star and obscuring part of its surface are one of the most 
promising methods of detection of extrasolar planets and other small size 
objects like brown dwarfs or small stars. When combined with precise 
spectroscopic observations, the method provides a unique opportunity to 
unambiguously determine all important parameters of a small companion like its 
size and mass. 

Detection of planetary transits requires, however, extremely accurate 
photometric data. For instance, the drop of brightness caused by a Jupiter 
size planet transiting across a solar-like star is of only 1\% order. Thus a 
millimagnitude accuracy of measurements is needed for detection. This is not 
easy to achieve from the ground. For firm detection and  to find the 
photometric orbit and orbital period observations of several (${\geq3}$) 
transits are necessary. On the other hand, when the photometric orbit is well 
known, only a few spectroscopic measurements at appropriate phases (around 
0.25 and 0.75) suffice to determine firm masses or at least constraints on the 
mass. It is often not realized that radial velocity measurements with accuracy 
of only 1--2~km/s are sufficient to distinguish between planetary and more 
massive object domains if no variation is found, provided that the star is not 
blended with an unresolved neighbor, physically related or optical. That limit 
can be achieved for very faint stars with the largest 8-m class modern 
telescopes with efficient spectrographs. Thus, the planetary search with 
transit method can be conducted even among objects located at large distances 
from the Sun, contrary to spectroscopic surveys limited to the solar 
neighborhood. 

Low probability of favorable orientation of the companion's orbit requires 
large number of stars to be monitored photometrically for detection of objects 
with transiting companions. Therefore a long term photometric program must be 
conducted for selection of candidates and determination of their photometric 
orbits. Several such programs have been undertaken during the last couple of 
years (Gilliland \etal 2000: 47~Tuc; Quirrenbach \etal 2000, Street \etal 
2002, PISCES project -- Mochejska \etal 2002: open clusters; EXPLORE project 
-- Yee \etal 2002: Galactic disk fields; STARE project -- Brown and 
Charbonneau 2000, VULCAN project -- Borucki \etal 2001: brighter stars) but 
only single transit-like light curves have been reported so far. It is worth 
noting that the only known object with planetary transiting companion --
HD~209458 (Henry \etal 2000 and Charbonneau \etal 2000) was discovered in the 
opposite way: after detection during the spectroscopic survey, the star was 
monitored photometrically for transits. 

Search for planetary and low luminosity transiting objects around main 
sequence stars became one of the highest priority goals of the third phase of 
the Optical Gravitational Lensing Experiment survey (OGLE-III). After the 
upgrade of detector with a wide field ${8192\times8192}$ pixel eight chip 
mosaic CCD camera and the software photometric data pipeline with algorithms 
based on the technique called ``image subtraction'' or ``difference image 
analysis'' (DIA) developed by Alard and Lupton (1998), Alard (2000) and 
Wo\'zniak (2000), the OGLE project became capable to monitor millions of stars 
with the long term accuracy of a few millimagnitudes for the brightest stars. 

The first ``planetary and low-luminosity object transit'' campaign was 
conducted by OGLE in June and July 2001 (Udalski \etal 2002ab). Three fields 
(${35\arcm\times35\arcm}$ each) of extremely high stellar density in the 
direction of the Galactic bulge were monitored continuously with the time 
resolution of about 12 minutes. Altogether 59 objects out of 52~000 Galactic 
disk stars with photometry better than 1.5\% were found to show small depth, 
flat-bottomed transits indicating presence of small size companions around 
these stars. For the vast majority of objects photometric elements could be 
derived as the individual transits were observed many times. Preliminary 
analysis of light curves indicated that several  companions can be 
Jupiter-sized, so they could be planets, brown dwarfs or late M-type dwarfs. 
Unfortunately, the photometry alone cannot unambiguously distinguish between 
these objects, as all of them may have the radii of the order of 0.1--0.2~\RS\ 
(1--2~$R_{\rm Jup}$). Therefore measurements of the radial velocity amplitude 
of the primaries are needed to determine the masses of transiting companions. 
Several of the OGLE-III transiting objects were followed-up spectroscopically 
with moderate resolution to better constrain sizes of the companions (Dreizler 
\etal 2002) as well as with high resolution spectroscopy. However, results of 
the latter have not been known at the moment of writing this paper. Apart from 
the final outcome of the first OGLE-III campaign, it clearly proved that the 
photometry of huge number of stars with high photometric accuracy necessary 
for transit detection can be obtained, and massive detection of low depth 
transits is feasible. 

The Galactic bulge line-of-sight is certainly not a perfect one for transit 
search. Luminous background of the Galactic bulge causes higher probability of 
the so called blending effect. Additional unresolvable light within the 
seeing disk of a star with a transiting object makes the depth of the transit 
shallower. In the worst case of very high blending, regular eclipsing star can 
mimic low amplitude transit. While in general the amount of blending can be 
modeled from the shape of transit or estimated from HST observations, it is 
desirable to lower the probability of blending by selection of fields in much 
less crowded areas like, for instance, the Galactic disk. 

In this paper we present results of the second ``planetary and low-luminosity 
object transit'' campaign conducted by OGLE-III in 
February--May 2002. Similarly to our first campaign three fields were 
monitored regularly with  high time resolution. These fields were located in 
the Carina regions of the Galactic disk at ${l\approx290\arcd}$. Analysis of 
the collected data led to discovery of more than 60 new objects with 
low-luminosity transiting companions. The sample contains many Jupiter-size 
objects -- very good candidates for extrasolar planets. Depth of some of the 
transits is at only several millimagnitude level. 

\newpage
Similarly to our first transit sample (Udalski \etal 2002ab) we decided to 
release the photometric data of our new candidates from the Carina fields to 
public domain so the follow-up spectroscopy could be made in short time scale 
by astronomers worldwide. Details and pointers to the OGLE {\sc Internet} 
archive can be found at the end of this paper. 

\Section{Observational Data}
Observations presented in this paper were collected with the 1.3-m Warsaw 
telescope at the Las Campanas Observatory, Chile (operated by the Carnegie 
Institution of  Washington), equipped with a wide field CCD mosaic camera. The 
camera consists of eight ${2048\times4096}$ pixel SITe ST002A detectors. The 
pixel size of each of the detectors is 15~$\mu$m giving the 0.26 arcsec/pixel 
scale at the focus  of the Warsaw telescope. Full field of view of the camera 
is about ${35\arcm\times35\arcm}$. The gain of each chip is adjusted to be 
about  1.3~e$^-$/ADU with the readout noise of about 6 to 9~e$^-$, depending 
on the chip. 
 
The photometric data were collected during 76 nights spanning 95 days starting 
from February 17, 2002. Three fields located in Carina regions of the Galactic 
disk were observed continuously with the time resolution of about 15 minutes. 
The fields were monitored up to 6 hours per night. Acronyms and equatorial 
coordinates of the fields are provided in Table~1. One of the fields, namely 
CAR100, was also sporadically observed before the main campaign. 

\MakeTable{lcc}{12.5cm}{Equatorial coordinates of transit fields}
{
\hline
\noalign{\vskip2pt}
\multicolumn{1}{c}{Field} & RA (J2000) & DEC (J2000)\\
\hline
\noalign{\vskip3pt}
CAR100  &  11\uph07\upm00\ups & $-61\arcd06\arcm30\arcs$ \\
CAR104  &  10\uph57\upm30\ups & $-61\arcd40\arcm00\arcs$ \\
CAR105  &  10\uph52\upm20\ups & $-61\arcd40\arcm00\arcs$ \\
\hline}

All observations were made in the {\it I}-band filter. We decided to increase 
the exposure time compared to the OGLE-III 2001 campaign to probe somewhat 
fainter stars, \ie in general of later spectral types. Therefore the exposure 
time of each image was set to 180 seconds, at the cost of somewhat smaller 
time resolution (about 15 minutes). Altogether about 1150 epochs were 
collected for each field during our 2002 campaign. The median seeing of the 
entire dataset was about $1\zdot\arcs2$. 

\Section{Data Reductions}
All collected images were preprocessed (de-biasing and flat-fielding) in real 
time with the standard OGLE-III data pipeline. Photometric reductions were 
performed off-line after the end of campaign when all frames were collected. 
Similarly to our 2001 campaign we applied the new OGLE-III photometric data 
pipeline based on the difference image analysis (DIA) method (Alard and Lupton 
1998, Alard 2000) and implementation of this method by Wo{\'z}niak (2000). For 
more details the reader is referred to Udalski \etal (2002a). We only note 
here, that the reference images were obtained by averaging of about 12 best 
images taken at the seeing of about 0\zdot\arcs8--0\zdot\arcs9. 

At this phase of the OGLE-III project observations are mainly focused on the 
variability. Therefore no calibration to the standard system of the images 
collected so far has been performed. No standard stars were observed during 
the 2002 campaign as well. Fortunately, part of the CAR100 field overlaps with 
CAR$\_$SC1, CAR$\_$SC2 and CAR$\_$SC3 fields observed and well calibrated 
during the OGLE-II phase. Based on the mean magnitudes of 13 transit 
candidates observed during both -- OGLE-II and OGLE-III phases we determined 
the mean shift of the magnitude scale between the OGLE-III magnitudes and 
OGLE-II calibrated data. This shift was applied not only to CAR100 but also to 
the remaining Carina fields. Our previous experience indicates that the error 
of the magnitude scale should not exceed 0.1--0.15~mag. Although OGLE-II 
photometry exists for 13 of our candidates, it should be noted that it is not 
suitable for search or even confirmation of transits because of much worse 
photometric quality obtained with the standard PSF fitting technique, much 
smaller number of observations and  non-appropriate sampling. 

Astrometric solution for the observed fields was performed in similar manner 
as for the 2001 campaign data (Udalski \etal 2002a), \ie by 
cross-identification of about 2000 brightest stars in each chip image with the 
Digitized Sky Survey images of the same part of the sky. Then the 
transformation between OGLE-III pixel grid and equatorial coordinates of the 
DSS (GSC) astrometric system was calculated.  This method was modified to 
obtain the initial transformation using WCSTools package (Mink 1997) and 
USNO-A2.0 Catalogue (Monet \etal 1998). The systematic error of the DSS 
astrometric solution can be up to about 0.7~arcsec, while the internal error 
of the transformation is about 0.2~arcsec. 

\Section{Search for Transits}
Before the transit search algorithm was applied to the collected data, a 
preselection procedure was performed. Similarly to the 2001 campaign we 
limited our search to stars with very precise photometry. We set the threshold 
of photometric accuracy at ${\leq15}$~mmag ({\it rms} from the entire time 
series). However, contrary to the 2001 campaign we did not make any limitation 
based on colors of preselected stars. In the Galactic bulge case that step was 
necessary to get rid of the Galactic bulge giants having similar magnitude as 
foreground Galactic disk stars. In the Carina Galactic disk fields the number 
of giants is very small, what can be deduced from the OGLE-II Carina fields 
color-magnitude diagrams, so they do not contaminate the sample in any 
significant way. Additionally, by lifting color constraints we do not miss, 
for instance, late K or M-type nearby objects which could also potentially 
host planets or other small companions. About 103~000 stars passed our ``good 
photometry'' cut. 

In the next step, all stars were subject to the transit search
algorithm. We  decided to run the BLS algorithm (Kov{\'a}cs, Zucker and
Mazeh 2002) which we find  to be very fast and efficient, based on our
experience from the 2001 campaign  (Udalski \etal 2002b). We used similar
parameters of the BLS algorithm as in  the 2001 campaign and limited our
search for transits to periods from 1.05 to  10~days. 

To our surprise, the BLS procedure run on our time series produced initially a 
huge number of artifacts, with transit-like light curves and similar pattern 
of periodicities. It was soon realized that photometry on a few nights was 
affected by the same variability pattern that folded to transit-like light 
curve shape and was triggered by the BLS algorithm. Most likely the photometry 
on these nights was affected by clouds or non-photometric conditions. After 
removing the data from these (four) nights the BLS procedure rerun on the 
entire dataset triggered much more reasonable number of candidates. 

The final list was prepared after a careful visual inspection of all light 
curves which passed the BLS algorithm. We left on the main list of candidates 
only those stars which we believe have a significant probability of being true 
transits. We removed a large number of small amplitude events caused by 
grazing eclipses  of regular stars of similar size and brightness (V-shaped 
eclipses) or somewhat deeper (${>0.1}$~mag) transits which occasionally passed 
our filters. However, we should stress that in the case of more noisy light 
curves it is not easy to distinguish between grazing eclipses and very 
non-central transits. Therefore, some of the stars on our list might be double 
stars, what can be easily verified in the future by spectroscopy. 
 
The final periods of our candidates were found after a careful examination of 
the eclipse light curve -- by minimizing dispersion during the eclipse phases 
that are very sensitive to the period changes. When only small number of transits was 
registered we always adopted the shortest period consistent with our remaining 
data. The formal accuracy of periods depends on the number and span of 
individual transits and it is of the order of $5\cdot10^{-4}$--$2\cdot10^{-4}P$. 

\Section{Results of 2002 Campaign}
Sixty two stars passed our filtering. We decided to keep on the list all stars 
with transiting companions when the transit depth was smaller than 0.08~mag. 
Such a limit corresponds to the companion size of $1.4~R_{\rm Jup}$ if the 
stellar radius is half of the solar. Although our search was limited to 
periods longer than 1.05 days, a few objects with shorter periods also entered 
the list. They were detected with the period of $2P$. 

Table~2 contains all basic data on our Carina objects with transiting 
companions. To preserve our notation from the OGLE-III 2001 campaign the first 
object in Table~2 is designated as OGLE-TR-60. In the subsequent columns of 
Table~2 the following data are provided: Identification, equatorial 
coordinates (J2000), orbital period, epoch of mid-eclipse, {\it I}-band 
magnitude outside transit, the depth of transit, number of transits observed 
($N_{\rm tr}$) and remarks. Accuracy of the magnitude scale is of about 
0.1--0.15 mag. OII abbreviation in the remarks column indicates that the 
object was also observed during the OGLE-II phase. 

Additionally, we present in Appendix the light curves and finding charts. For 
each object the full light curve and close-up around the transit are shown. 
Please note that the magnitude scale changes in the close-up windows, 
depending on brightness, noise and transit depth. The finding chart is a 
${60\arcs\times60\arcs}$ subframe of the {\it I}-band reference image centered 
on the star. The star is marked by a white cross. North is up and East to the 
left in these images. 
\MakeTable{l@{\hspace{6pt}}
c@{\hspace{5pt}}c@{\hspace{6pt}}c@{\hspace{5pt}}r@{\hspace{4pt}}
c@{\hspace{4pt}}c@{\hspace{4pt}}r@{\hspace{4pt}}l}{12.5cm}
{OGLE-III planetary and low luminosity object transits} 
{\hline
\noalign{\vskip4pt} 
\multicolumn{1}{c}{Name} & RA (J2000)  & DEC (J2000) &   $P$    &
\multicolumn{1}{c}{$T_0$}& $I$    &$\Delta I$  &  \multicolumn{1}{c}{$N_{\rm tr}$}&Rem.\\
           &             &             & [days]   & --2452000     &[mag] &[mag]  &\\ 
\noalign{\vskip4pt}
\hline
\noalign{\vskip4pt}
OGLE-TR-60 & 11\uph08\upm37\zdot\ups25 & $-61\arcd20\arcm16\zdot\arcs7$ & 2.30890 &  75.77838 & 14.60 & 0.016 & 11 & OII \\
OGLE-TR-61 & 11\uph08\upm41\zdot\ups13 & $-61\arcd07\arcm58\zdot\arcs5$ & 4.26800 &  76.62809 & 16.26 & 0.030 &  8 & OII \\
OGLE-TR-62 & 11\uph08\upm37\zdot\ups24 & $-61\arcd10\arcm45\zdot\arcs3$ & 2.60119 &  77.66751 & 15.91 & 0.038 & 10 & OII \\
OGLE-TR-63 & 11\uph08\upm58\zdot\ups01 & $-61\arcd01\arcm29\zdot\arcs4$ & 1.06698 &  74.24553 & 15.75 & 0.011 & 12 & OII \\
OGLE-TR-64 & 11\uph07\upm50\zdot\ups90 & $-61\arcd05\arcm39\zdot\arcs9$ & 2.71740 &  75.85150 & 16.17 & 0.022 &  7 & OII \\
OGLE-TR-65 & 11\uph07\upm53\zdot\ups46 & $-61\arcd04\arcm18\zdot\arcs2$ & 0.86013 &  76.31928 & 15.94 & 0.034 & 18 & OII \\
OGLE-TR-66 & 11\uph07\upm04\zdot\ups18 & $-60\arcd54\arcm21\zdot\arcs0$ & 3.51407 &  74.42473 & 15.18 & 0.053 &  6 & \\
OGLE-TR-67 & 11\uph08\upm57\zdot\ups92 & $-60\arcd52\arcm58\zdot\arcs9$ & 5.27980 &  78.76236 & 16.40 & 0.053 &  5 & OII \\
OGLE-TR-68 & 11\uph05\upm48\zdot\ups85 & $-60\arcd54\arcm50\zdot\arcs6$ & 1.28870 &  73.10520 & 16.79 & 0.030 & 12 & \\
OGLE-TR-69 & 11\uph06\upm06\zdot\ups40 & $-60\arcd56\arcm19\zdot\arcs9$ & 2.33708 &  75.19708 & 16.55 & 0.038 &  5 & \\
OGLE-TR-70 & 11\uph05\upm12\zdot\ups33 & $-61\arcd14\arcm00\zdot\arcs2$ & 8.04060 &  77.22138 & 16.89 & 0.053 &  4 & OII \\
OGLE-TR-71 & 11\uph06\upm52\zdot\ups67 & $-61\arcd14\arcm16\zdot\arcs4$ & 4.18760 &  76.34670 & 16.38 & 0.022 &  5 & OII \\
OGLE-TR-72 & 11\uph05\upm59\zdot\ups06 & $-61\arcd10\arcm08\zdot\arcs1$ & 6.85400 &  77.39865 & 16.44 & 0.048 &  4 & OII \\
OGLE-TR-73 & 11\uph05\upm33\zdot\ups61 & $-61\arcd08\arcm45\zdot\arcs3$ & 1.58105 &  73.42506 & 16.99 & 0.034 &  9 & OII \\
OGLE-TR-74 & 11\uph06\upm10\zdot\ups71 & $-61\arcd14\arcm52\zdot\arcs7$ & 1.58511 &  76.33895 & 15.87 & 0.030 & 11 & OII \\
OGLE-TR-75 & 11\uph06\upm37\zdot\ups91 & $-61\arcd19\arcm15\zdot\arcs5$ & 2.64270 &  77.35886 & 16.96 & 0.034 &  8 & OII \\
OGLE-TR-76 & 10\uph58\upm41\zdot\ups90 & $-61\arcd53\arcm12\zdot\arcs3$ & 2.12678 & 323.54517 & 13.76 & 0.022 &  6 & \\
OGLE-TR-77 & 10\uph58\upm02\zdot\ups03 & $-61\arcd49\arcm50\zdot\arcs9$ & 5.45550 & 326.45043 & 16.12 & 0.022 &  4 & \\
OGLE-TR-78 & 10\uph59\upm41\zdot\ups62 & $-61\arcd55\arcm15\zdot\arcs0$ & 5.32038 & 328.81199 & 15.32 & 0.030 &  4 & \\
\hline}

\setcounter{table}{1}
\MakeTableSep{l@{\hspace{6pt}}
c@{\hspace{5pt}}c@{\hspace{6pt}}c@{\hspace{5pt}}r@{\hspace{4pt}}
c@{\hspace{4pt}}c@{\hspace{4pt}}r@{\hspace{4pt}}l}{12.5cm}
{Concluded} 
{\hline
\noalign{\vskip4pt} 
\multicolumn{1}{c}{Name} & RA (J2000)  & DEC (J2000) &   $P$    &
\multicolumn{1}{c}{$T_0$}& $I$    &$\Delta I$  &  \multicolumn{1}{c}{$N_{\rm tr}$}&Rem.\\
           &             &             & [days]   & --2452000     &[mag] &[mag]  &\\ 
\noalign{\vskip4pt}
\hline
\noalign{\vskip4pt}
OGLE-TR-79 & 10\uph59\upm35\zdot\ups54 & $-61\arcd56\arcm59\zdot\arcs2$ & 1.32452 & 324.28819 & 15.28 & 0.030 & 13 & \\
OGLE-TR-80 & 10\uph57\upm54\zdot\ups36 & $-61\arcd42\arcm02\zdot\arcs5$ & 1.80730 & 325.49707 & 16.50 & 0.016 & 12 & \\
OGLE-TR-81 & 10\uph59\upm26\zdot\ups49 & $-61\arcd36\arcm49\zdot\arcs0$ & 3.21650 & 323.10755 & 15.41 & 0.022 &  6 & \\
OGLE-TR-82 & 10\uph58\upm03\zdot\ups07 & $-61\arcd34\arcm25\zdot\arcs8$ & 0.76416 & 323.08758 & 16.30 & 0.034 & 22 & \\
OGLE-TR-83 & 10\uph57\upm42\zdot\ups48 & $-61\arcd36\arcm23\zdot\arcs3$ & 1.59920 & 323.20108 & 14.87 & 0.016 & 12 & \\
OGLE-TR-84 & 10\uph59\upm00\zdot\ups00 & $-61\arcd34\arcm43\zdot\arcs0$ & 3.11300 & 324.98303 & 16.69 & 0.059 &  6 & \\
OGLE-TR-85 & 10\uph59\upm00\zdot\ups18 & $-61\arcd37\arcm41\zdot\arcs1$ & 2.11460 & 324.44012 & 15.45 & 0.048 & 12 & \\
OGLE-TR-86 & 10\uph58\upm19\zdot\ups30 & $-61\arcd29\arcm27\zdot\arcs3$ & 2.77700 & 323.92937 & 16.32 & 0.065 &  7 & \\
OGLE-TR-87 & 10\uph59\upm39\zdot\ups33 & $-61\arcd24\arcm07\zdot\arcs3$ & 6.60672 & 332.12239 & 16.32 & 0.059 &  3 & \\
OGLE-TR-88 & 10\uph59\upm22\zdot\ups23 & $-61\arcd25\arcm21\zdot\arcs0$ & 1.25012 & 323.58871 & 14.58 & 0.034 & 15 & \\
OGLE-TR-89 & 10\uph56\upm11\zdot\ups27 & $-61\arcd29\arcm55\zdot\arcs4$ & 2.28990 & 323.00793 & 15.78 & 0.013 &  5 & \\
OGLE-TR-90 & 10\uph56\upm36\zdot\ups63 & $-61\arcd28\arcm46\zdot\arcs5$ & 1.04155 & 322.67722 & 16.44 & 0.022 & 15 & \\
OGLE-TR-91 & 10\uph57\upm31\zdot\ups20 & $-61\arcd27\arcm21\zdot\arcs7$ & 1.57900 & 324.30930 & 15.23 & 0.043 &  9 & \\
OGLE-TR-92 & 10\uph57\upm23\zdot\ups43 & $-61\arcd26\arcm45\zdot\arcs4$ & 0.97810 & 322.96414 & 16.50 & 0.038 & 20 & \\
OGLE-TR-93 & 10\uph55\upm20\zdot\ups36 & $-61\arcd24\arcm59\zdot\arcs4$ & 2.20674 & 324.91426 & 15.20 & 0.019 & 12 & \\
OGLE-TR-94 & 10\uph55\upm48\zdot\ups86 & $-61\arcd28\arcm44\zdot\arcs5$ & 3.09222 & 327.38138 & 14.32 & 0.043 &  6 & \\
OGLE-TR-95 & 10\uph55\upm19\zdot\ups38 & $-61\arcd32\arcm12\zdot\arcs0$ & 1.39358 & 325.30146 & 16.36 & 0.019 & 14 & \\
OGLE-TR-96 & 10\uph56\upm33\zdot\ups99 & $-61\arcd37\arcm10\zdot\arcs5$ & 3.20820 & 323.51865 & 14.90 & 0.043 &  6 & \\
OGLE-TR-97 & 10\uph55\upm17\zdot\ups94 & $-61\arcd54\arcm35\zdot\arcs7$ & 0.56765 & 322.83189 & 15.51 & 0.016 & 25 & \\
OGLE-TR-98 & 10\uph56\upm51\zdot\ups77 & $-61\arcd56\arcm15\zdot\arcs0$ & 6.39800 & 327.77953 & 16.64 & 0.034 &  5 & \\
OGLE-TR-99 & 10\uph55\upm12\zdot\ups80 & $-61\arcd54\arcm54\zdot\arcs8$ & 1.10280 & 323.36419 & 16.47 & 0.034 & 16 & \\
OGLE-TR-100& 10\uph52\upm56\zdot\ups91 & $-61\arcd50\arcm54\zdot\arcs9$ & 0.82670 & 323.32754 & 14.88 & 0.019 & 20 & \\
OGLE-TR-101& 10\uph52\upm58\zdot\ups59 & $-61\arcd51\arcm43\zdot\arcs1$ & 2.36180 & 324.23024 & 16.69 & 0.038 &  8 & \\
OGLE-TR-102& 10\uph53\upm29\zdot\ups65 & $-61\arcd47\arcm37\zdot\arcs2$ & 3.09790 & 323.79252 & 13.84 & 0.019 &  5 & \\
OGLE-TR-103& 10\uph53\upm33\zdot\ups53 & $-61\arcd47\arcm04\zdot\arcs3$ & 8.21690 & 324.30267 & 16.69 & 0.048 &  4 & \\
OGLE-TR-104& 10\uph53\upm27\zdot\ups04 & $-61\arcd43\arcm20\zdot\arcs3$ & 6.06800 & 328.02979 & 17.10 & 0.053 &  2 & \\
OGLE-TR-105& 10\uph52\upm24\zdot\ups07 & $-61\arcd31\arcm09\zdot\arcs4$ & 3.05810 & 324.37986 & 16.16 & 0.026 &  3 & \\
OGLE-TR-106& 10\uph53\upm51\zdot\ups23 & $-61\arcd34\arcm13\zdot\arcs2$ & 2.53585 & 324.78332 & 16.53 & 0.022 &  6 & \\
OGLE-TR-107& 10\uph54\upm23\zdot\ups58 & $-61\arcd37\arcm21\zdot\arcs1$ & 3.18980 & 323.55936 & 16.66 & 0.053 &  7 & \\
OGLE-TR-108& 10\uph53\upm12\zdot\ups65 & $-61\arcd30\arcm18\zdot\arcs7$ & 4.18590 & 325.78009 & 17.28 & 0.048 &  3 & \\
OGLE-TR-109& 10\uph53\upm40\zdot\ups73 & $-61\arcd25\arcm14\zdot\arcs8$ & 0.58909 & 323.74379 & 14.99 & 0.008 & 24 & \\
OGLE-TR-110& 10\uph52\upm28\zdot\ups37 & $-61\arcd29\arcm31\zdot\arcs8$ & 2.84857 & 326.36185 & 16.15 & 0.026 &  6 & \\
OGLE-TR-111& 10\uph53\upm17\zdot\ups91 & $-61\arcd24\arcm20\zdot\arcs3$ & 4.01610 & 330.44687 & 15.55 & 0.019 &  9 & \\
OGLE-TR-112& 10\uph52\upm46\zdot\ups46 & $-61\arcd23\arcm17\zdot\arcs7$ & 3.87900 & 327.53260 & 13.64 & 0.016 &  8 & \\
OGLE-TR-113& 10\uph52\upm24\zdot\ups40 & $-61\arcd26\arcm48\zdot\arcs5$ & 1.43250 & 324.36394 & 14.42 & 0.030 & 10 & \\
OGLE-TR-114& 10\uph52\upm20\zdot\ups79 & $-61\arcd29\arcm45\zdot\arcs2$ & 1.71213 & 323.24893 & 15.76 & 0.026 &  5 & \\
OGLE-TR-115& 10\uph50\upm20\zdot\ups50 & $-61\arcd28\arcm34\zdot\arcs4$ & 8.34670 & 329.94514 & 16.66 & 0.059 &  3 & \\
OGLE-TR-116& 10\uph50\upm24\zdot\ups79 & $-61\arcd26\arcm12\zdot\arcs2$ & 6.06430 & 324.21555 & 14.90 & 0.077 &  5 & \\
OGLE-TR-117& 10\uph51\upm40\zdot\ups48 & $-61\arcd34\arcm15\zdot\arcs7$ & 5.02260 & 325.29544 & 16.71 & 0.030 &  5 & \\
OGLE-TR-118& 10\uph51\upm32\zdot\ups10 & $-61\arcd48\arcm08\zdot\arcs3$ & 1.86150 & 326.26908 & 17.07 & 0.019 &  7 & \\
OGLE-TR-119& 10\uph51\upm58\zdot\ups75 & $-61\arcd41\arcm20\zdot\arcs5$ & 5.28260 & 323.67601 & 14.29 & 0.038 &  7 & \\
OGLE-TR-120& 10\uph51\upm09\zdot\ups34 & $-61\arcd43\arcm11\zdot\arcs3$ & 9.16590 & 331.49765 & 16.23 & 0.077 &  4 & \\
OGLE-TR-121& 10\uph50\upm36\zdot\ups41 & $-61\arcd40\arcm37\zdot\arcs2$ & 3.23210 & 325.68889 & 15.86 & 0.071 &  6 & \\
\hline}

\Section{Discussion}
Sixty two new objects with transiting companions were discovered during the 
second ``planetary and low-luminosity object transit'' observational campaign 
conducted by OGLE-III in 2002 increasing the total number of transiting 
objects found by OGLE to 121. For each new object several individual transits 
were detected and determination of the photometric ephemerides was possible. 

Transits can be caused by extrasolar planets or brown dwarfs or small late 
M-type dwarfs. To distinguish between these possibilities, radial velocity 
measurements are necessary and we hope they will be obtained in the near 
future. However, one should remember that blending of a regular totally 
eclipsing star with a close optical or physically related (wide binary system) 
unresolvable neighbor can produce transit-like light curve. Therefore some of 
our candidates listed in Table~2 can actually be faked transits caused by 
blending effect. High resolution spectroscopy should also clarify this 
problem. 

Photometric data alone allow to draw conclusions only on sizes of transiting 
objects. Unfortunately, without any additional information on the radius of the 
primary it is not possible to obtain actual size of the companion in a system 
with transits when the errors of individual observations are comparable to the 
transit depth. Due to well known degeneracy between radii of the host star and 
companion, $R_s$, $R_c$, inclination, $i$, and limb darkening, $u$, similar 
quality photometric solutions can be obtained for different inclinations of 
the orbit and radii of components (in the {\it I}-band the transit light curve 
is practically insensitive to the limb darkening parameter $u$). Such 
additional information can come from moderate resolution spectroscopy suitable 
for spectral classification (Dreizler \etal 2002) or, in principle, from colors 
of host stars. Unfortunately, the latter are of no use in the case of the pencil 
beam survey of the Galactic disk, because usually the significant and  unknown 
interstellar extinction makes dereddening of individual stars practically 
impossible. 

Because the spectral types of stars from our sample are not known and the 
primary radius cannot be constrained, only the lower limit on the size of the 
companion can be calculated assuming that the transit is central, \ie 
${i=90\arcd}$. The corresponding radius of the primary is also the lower 
limit.  Table~3 lists lower limits of the components radii calculated using 
formulae provided by Sackett (1999) under the assumption that the mass of the 
primary is equal to ${M_s=1~\MS}$. It should be remembered that the values in 
Table~3 scale as $M^{1/3}_s$. Details of the modeling of transit light curve are 
given in Udalski \etal (2002a). 

Solid line in the close-up windows in Appendix shows the transit model light 
curve calculated for the central passage. As it can be seen, in some cases the 
fit is not satisfactory indicating the inclination smaller than $90\arcd$. 
However, in most cases the central passage fit is practically 
indistinguishable from others so at this stage it is impossible to derive 
other values than the lower limits of radii provided in Table~3. 

The limits of radii of transiting companions may be used for the first 
preselection of our 2002 transit sample. For several objects the lower limits of 
radii of the companions are larger than 0.25~\RS. These systems almost 
certainly contain M-type dwarf companions. The remaining objects can be either 
planets or brown dwarfs or late M-type dwarfs. The smaller size of the 
companion -- the larger probability that the companion is an extrasolar 
planet. But, on the other hand, one should always remember that the figures in 
Table~3 are only the lower limits and the real radius of the companion in any
given case might be much larger. 

\vspace*{-9pt}
\renewcommand{\arraystretch}{0.9}
\renewcommand{\TableFont}{\footnotesize}
\MakeTable{lccclcc}{12.5cm}{Dimensions of stars and companions for central
passage ($M_s=1~\MS$)}
{\cline{1-3}\cline{5-7}
\noalign{\vskip3pt}
Name       & $R_s$   & $R_c$ & $\phantom{xxxxxxx}$ & Name & $R_s$ & $R_c$\\
&[\RS]&[\RS]&$\phantom{xxxxx}$ &&[\RS]&[\RS]\\
\noalign{\vskip3pt}
\cline{1-3}\cline{5-7}
\noalign{\vskip3pt}
OGLE-TR-60 & 1.60 & 0.176&  & OGLE-TR-91 & 1.43 & 0.257 \\
OGLE-TR-61 & 3.46 & 0.519&  & OGLE-TR-92 & 0.90 & 0.153 \\
OGLE-TR-62 & 1.66 & 0.282&  & OGLE-TR-93 & 1.70 & 0.204 \\
OGLE-TR-63 & 1.16 & 0.104&  & OGLE-TR-94 & 0.89 & 0.161 \\
OGLE-TR-64 & 1.14 & 0.148&  & OGLE-TR-95 & 1.25 & 0.150 \\
OGLE-TR-65 & 1.00 & 0.160&  & OGLE-TR-96 & 1.09 & 0.196 \\
OGLE-TR-66 & 1.08 & 0.216&  & OGLE-TR-97 & 1.11 & 0.122 \\
OGLE-TR-67 & 1.82 & 0.364&  & OGLE-TR-98 & 1.42 & 0.227 \\
OGLE-TR-68 & 1.08 & 0.162&  & OGLE-TR-99 & 0.98 & 0.156 \\
OGLE-TR-69 & 1.28 & 0.217&  & OGLE-TR-100& 0.98 & 0.117 \\
OGLE-TR-70 & 0.34 & 0.067&  & OGLE-TR-101& 0.70 & 0.120 \\
OGLE-TR-71 & 0.75 & 0.097&  & OGLE-TR-102& 1.12 & 0.134 \\
OGLE-TR-72 & 1.08 & 0.206&  & OGLE-TR-103& 1.36 & 0.258 \\
OGLE-TR-73 & 0.84 & 0.134&  & OGLE-TR-104& 1.84 & 0.368 \\
OGLE-TR-74 & 0.95 & 0.143&  & OGLE-TR-105& 1.17 & 0.163 \\
OGLE-TR-75 & 1.55 & 0.247&  & OGLE-TR-106& 1.22 & 0.159 \\
OGLE-TR-76 & 0.85 & 0.110&  & OGLE-TR-107& 1.11 & 0.222 \\
OGLE-TR-77 & 2.25 & 0.292&  & OGLE-TR-108& 1.37 & 0.259 \\
OGLE-TR-78 & 0.97 & 0.146&  & OGLE-TR-109& 1.23 & 0.099 \\
OGLE-TR-79 & 1.27 & 0.190&  & OGLE-TR-110& 0.98 & 0.138 \\
OGLE-TR-80 & 1.56 & 0.172&  & OGLE-TR-111& 0.90 & 0.108 \\
OGLE-TR-81 & 2.18 & 0.284&  & OGLE-TR-112& 3.05 & 0.336 \\
OGLE-TR-82 & 0.76 & 0.122&  & OGLE-TR-113& 0.86 & 0.129 \\
OGLE-TR-83 & 0.99 & 0.109&  & OGLE-TR-114& 0.97 & 0.136 \\
OGLE-TR-84 & 1.28 & 0.269&  & OGLE-TR-115& 0.66 & 0.139 \\
OGLE-TR-85 & 1.28 & 0.244&  & OGLE-TR-116& 1.25 & 0.300 \\
OGLE-TR-86 & 0.78 & 0.172&  & OGLE-TR-117& 1.76 & 0.264 \\
OGLE-TR-87 & 1.26 & 0.264&  & OGLE-TR-118& 2.12 & 0.233 \\
OGLE-TR-88 & 1.28 & 0.206&  & OGLE-TR-119& 1.82 & 0.310 \\
OGLE-TR-89 & 0.96 & 0.096&  & OGLE-TR-120& 1.08 & 0.260 \\
OGLE-TR-90 & 0.68 & 0.088&  & OGLE-TR-121& 1.02 & 0.235 \\
\cline{1-3}\cline{5-7}}

One can notice that the number of objects with the limits on the companion 
smaller than ${1.6~R_{\rm Jup}}$ is large in our 2002 sample (26) -- much 
larger than on the 2001 object list (11). Also the limits on radii of primary 
stars are smaller than the values of those discovered during the 2001 
campaign. This is very likely because of the deeper range of our 2002 survey 
(longer exposures) so we probe on average later spectral type stars than 
during the 2001 season. If so, the probability of detection of extrasolar 
planets in our 2002 sample should be larger. Additionally, we detected many 
more objects with the transit depth of only several millimagnitudes compared 
to the 2001 campaign. 

The list of the promising planetary transit candidates among our objects
is  quite long so we enumerate only a few: OGLE-TR-71, OGLE-TR-113,  
OGLE-TR-111, OGLE-TR-90, OGLE-TR-89, OGLE-TR-101, OGLE-TR-100. We would
also like to draw attention to a group of about ten objects with 
orbital periods close to or smaller than one day. Short distance from
the host  star makes the transits long in phase, but the limits from
Table~3 suggest  solar type primaries. Very small depth of transits
indicates that companions  are well in the planetary range of sizes.
Unfortunately, in many cases the  observational scatter is somewhat too
high to definitively recognize the  transit light curve shape, so some
of them might be grazing eclipses. If confirmed by future follow-up
spectroscopy as planetary systems, the companions in these objects
would be the shortest orbital period extrasolar  planets providing
important constraints on the modeling of the origin and  evolution of
planetary systems. 

It is worth noting that not only extrasolar planetary cases are of great 
astronomical importance. Discovery of brown dwarfs among our candidates would 
contradict the existence of the so called ``brown dwarf desert'' (lack of 
brown dwarfs at small orbits in binary systems) and allow  for the first time 
to precisely measure their masses and sizes. Also determination of masses and 
sizes of late M-type dwarfs is extremely important as the mass-radius relation 
of the lower part of the main sequence is poorly known. Therefore we strongly 
encourage astronomers worldwide to make follow-up observations. During the 
next seasons (2003 and 2004) most of our photometric ephemerides should be 
accurate enough that only a few observations per star should suffice to 
determine precise masses. 

\Section{Data Availability}
The photometric data on the objects with transiting companions discovered 
during the 2002 OGLE-III campaign are available in the electronic form from 
the OGLE archive: 
\vskip3pt
\centerline{\it http://www.astrouw.edu.pl/\~{}ogle} 
\vskip3pt
\centerline{\it ftp://ftp.astrouw.edu.pl/ogle/ogle3/transits/carina$\_$2002}
\vskip3pt
\noindent
or its US mirror

\centerline{\it http://bulge.princeton.edu/\~{}ogle}
\vskip3pt
\centerline{\it ftp://bulge.princeton.edu/ogle/ogle3/transits/carina$\_$2002}

\vskip20pt
\Acknow{We would like to thank Prof.\ B.\ Paczy{\'n}ski for many interesting 
discussions and comments. The paper was partly supported by the  Polish KBN 
grant BST to Warsaw University Observatory. Partial support to the OGLE 
project was provided with the NSF grant AST-0204908 and NASA grant 
NAG5-12212 to B.~Paczy\'nski. We acknowledge usage of the Digitized Sky Survey 
which was produced at the Space Telescope Science Institute based on 
photographic data obtained using the UK Schmidt Telescope, operated by the 
Royal Observatory Edinburgh.}

\end{document}